\documentclass[envcountsame,12pt]{article}
\usepackage{amssymb}

\def\To{\Rightarrow}

\def\Tm{\mbox{\it Tm}}
\def\Fm{\mbox{\it Fm}}
\def\Expr{\mbox{\it Expr}}
\def\mod{\,\mbox{\it mod}\,}
\def\Jref{{\sf J}_{\mbox{\scriptsize\it ref}}}

\newtheorem{theo}{Theorem}
\newtheorem{lem}[theo]{Lemma}
\newtheorem{cor}[theo]{Corollary}
\newtheorem{defin}[theo]{Definition}

\def\proof{\par\smallskip\noindent {\it Proof. }}

\def\qed{\hfill$\lhd$\par\medskip}

\title{On the sharpness and the injective property of basic justification models}
\author{Vladimir N.~Krupski \\
{\small Faculty of Mechanics and Mathematics,}\\ {\small Lomonosov Moscow State
University, Moscow 119992, Russia}\\ {\small\tt krupski@lpcs.math.msu.su}}

\date{}

\begin{document}
\maketitle

\begin{abstract}
Justification Awareness Models, \ JAMs, were proposed by S.~Artemov as a tool for
modelling epistemic scenarios like Russel's Prime Minister example. It was demonstrated
that the sharpness and the injective property of a model play essential role in the
epistemic usage of JAMs. The problem to axiomatize these properties using the
propositional justification language was left opened. We propose the solution and define
a decidable justification logic $\Jref$ that is sound and complete with respect to the
class of all sharp injective justification models.
\end{abstract}

\section{Introduction}
Justification Awareness Models ({\it JAM}) were introduced in \cite{ArtEMJ} as a
flexible tool for modelling epistemic scenarios like Russel's Prime Minister
example.\footnote[1]{In \cite{ArtEMJ} they were referred as {\it JEM}s, Justification
Epistemic Models. Later the terminology was changed.} A {\it JAM} consists of a basic
model for justification logic ${\sf J}^{-}$ (see \cite{ArtLJ}), supplied with the means
to distinguish acceptable (i.e. meaningful) and knowledge-producing justifications.

In this paper we consider the first component. It is referred in \cite{ArtEMJ} as {\em a
basic justification model}. The language of the model extends the usual propositional
language by new atoms, {\em justification assertions}, of the form $t\!:\!F$ with the
intended meaning ``$t$ is a justification of $F$''. Justifications are terms built from
atomic ones by a binary operation $\cdot$ (application) that reflects logical reasonings
via {\em Modus ponens} rule, so the following property is assumed:
\begin{equation}\label{ApplicationAx}
s\!\!:\!(F\to G)\to(t\!\!:\!F\to [s\!\cdot\! t]\!\!:\!G).
\end{equation}

Justification logic ${\sf J}^{-}$ is the extension of the classical propositional logic
by Application axiom (\ref{ApplicationAx}) and a basic justification model (up to some
details of the formulation, see Section \ref{BJM}) corresponds to a single world in the
constructive canonical model for ${\sf J}^{-}$. In such a model a justification $t$
denotes the set of formulas justified by $t$ and the justification assertion $t\!:\!F$
means that $F$ is a member of this set. The application $\cdot$ denotes a binary
operation on sets of formulas that satisfied the condition (\ref{ApplicationAx}).

The epistemic usage of {\it JAM}s involves the detailed analysis of the term structure
of a model. The following properties of a model, the {\em sharpness} and the {\em
injective property}, are pointed out in \cite{ArtEMJ}  as essential.

Sharpness. Consider a model with some true justification assertion of the form
$[s\!\cdot\! t]\!\!:\!G$. It is a claim that $G$ follows by logical reasoning using {\em
Modus ponens} rule from some facts already justified by $s$ and $t$ respectively. One
should treat it as nonsense when there is no such facts. The sharpness condition
eliminates this possibility. It requires that application should be interpreted by the
following operation on sets of formulas:
$$
S\rhd T=\{G\mid F\to G\in S\mbox{ and }F\in T\mbox{ for some }F\}.
$$
So, in a sharp model the application means application of {\em Modus ponens} rule and
nothing more.

Injective justifications. A model is injective if for every justification $t$ there
exists at most one formula that is justified by $t$. This requirement admits the
treatment of justifications as objects, not only as parts of justification assertions.
The decision whether a justification $t$ is meaningful or knowledge-producing can be
made on the base of the analysis of $t$ itself and does not depend on the context where
it is used. The justified statement $v_{t}$ can be restored from it, so for meaningful
$t$ the justification assertion $t\!:\!F$ implies $v_{t}=F$ and $t\!:\!v_{t}$.

Justification logic ${\sf J}^{-}$ is sound and complete with respect to the class of all
basic justification models (see \cite{ArtLJ},\cite{ArtEMJ}). How to axiomatize the class
of all basic justification models that are sharp and injective? This question was stated
as an open problem in \cite{ArtEMJ}. We provide the solution.

The key idea is to distinguish between the language of a model and the language of the
logic. Both of them are justification languages but in the first one atoms are treated
as constants whereas in the second one they are syntactical variables that admit
substitution. An interpretation of the logical language in a model is an infinite
substitution that replaces syntactical variables by corresponding expressions of the
language of the model, the translation need not be injective. This approach gives the
possibility to axiomatize the injective property of a model via Unification axioms (see
\cite{ArtStr93TR},\cite{Kru},\cite{KruAPAL} where they are used for axiomatization of
the single-conclusion property of arithmetical proof predicates).

In the presence of Unification axioms the sharpness property can be expressed using
reference constructions $v_{t}$.  We add them to the logical language. Reference
constructions in the justification language were considered in
\cite{KruJLC},\cite{KruTCS} where the general technique was developed and used in the
context of Logic of Proofs. We simplify the exposition and adjust it to the case of
${\sf J}^{-}$ and the particular reference construction ``the judgement justified by
$t$''. As a result we obtain a decidable justification logic $\Jref$ and prove that it
is sound and complete with respect to the class of all sharp injective basic
justification models.

\section{Preliminaries}
\subsection{Basic justification models (cf. \cite{ArtEMJ})} \label{BJM}
Let $P^0$ (atomic propositions) and $J^0$ (atomic justifications) be disjoint countable
sets of identifiers. The justification language $L(P^0,J^0)$ has two sorts of
expressions
--- justification terms ($Tm^0$) and formulas ($\Fm^0$), defined by the following
grammar:
$$
\Tm^0 ::= J^0\mid \Tm^0\cdot \Tm^0, \qquad \Fm^0 ::= \bot\mid P^0\mid \Fm^0\to \Fm^0
\mid \Tm^0\!:\Fm^0.
$$

A {\em basic justification model} is defined in \cite{ArtEMJ} as a pair $\langle
L(P^0,J^0), *\rangle$ where $*$ is an interpretation that consists of two parts, $*\!:
\Fm^0\to \{0,1\}$,  $*\!: \Tm^0\to 2^{\Fm^0}$. It has the following properties:
$$
\bot^*=0, \qquad (F\to G)^*=1 \Leftrightarrow (F^*=0 \mbox{ or }G^*=1),
$$
$$
(t\!\!:\!F)^*=1 \Leftrightarrow F\in t^*, \qquad s^*\rhd t^*\subseteq (s\cdot t)^*.
$$

The class of all basic models can be axiomatized by the system ${\sf J}^{-}$ (see
\cite{ArtLJ}) which is asserted in \cite{ArtEMJ} to be the base system of justification
epistemic logic. Basic models correspond to possible worlds in the canonical model of
${\sf J}^{-}$, so ${\sf J}^{-}$ is sound and complete with respect to this semantics
(see \cite{ArtEMJ}).

A basic model is called {\em sharp} when $s^*\rhd t^* = (s\cdot t)^*$ for all $t,s\in
\Tm^0$. It is {\em injective} if for all $t\in \Tm^0$ the set $t^*$ contains no more
than one formula. These properties of a model become essential when we analyze the term
structure of justifications in more details. Injective justifications can be used as
pointers (see \cite{Kru}, \cite{KruJLC}, \cite{KruTCS} for details). Below we exploit
this ability in order to axiomatize the sharpness property.

\subsection{Unification}
We remind the unification technique developed in \cite{KruJLC}, \cite{KruTCS}. Let
$P=\{p_0,p_1,\ldots\}$ and $J=\{x_0,x_1,\ldots\}$ be sets of syntactical (first-order)
variables of two sorts. The language $L^{v}(P,J)$ is the extension of $L(P,J)$ by the
additional second-order function variable $v$ of type $Tm\to Fm$. It is defined by the
grammar
$$
\Tm ::= J\mid \Tm\cdot \Tm, \qquad \Fm ::= \bot\mid P\mid \Fm\to \Fm \mid \Tm\!:\Fm \mid
v(\Tm),
$$
so expressions of the form $v(t)$ are additional first-order variables indexed by terms,
$v_t=v(t)$. Below we use this notation for better readability.\footnote[2]{In
\cite{KruJLC}, \cite{KruTCS} these variables  are called {\em reference constructions}.
In the context of Single-Conclusion Logic of Proofs they represent syntactical
operations that restore some parts of a formula given its proof. It will be seen that
$v$ corresponds to the proof goal operation that extracts a formula from its proof.}

Members of $\Expr=\Tm\cup \Fm$ will be considered as terms in the signature $\Omega
=\{\bot ,\rightarrow ,: ,\cdot \}$ and will be called {\em expressions}. In this context
{\em a substitution} is a sort preserving homomorphism of free term algebras of
signature $\Omega$, i.e. a function on $Expr$ that maps terms into terms, formulas into
formulas and commutes with symbols from $\Omega$.

We admit infinite substitutions too. A substitution $\theta$ is completely defined by
its values on atomic expressions from the set $Var=J\cup P \cup v(\Tm)$. Let
$$
Dom(\theta) =\{z\in Var\mid z\theta\not=z\},\quad Var(\theta)= \bigcup_{z\in
Dom(\theta)} Var(z\theta)\cup Dom(\theta),
$$
where $Var(e)$ denotes the set of all $z\in Var$ that occur in $e\in \Expr$.

A substitution $\theta$ is called {\em comprehensive}  if $t_1\theta =t_2\theta$ implies
$v_{t_1}\theta =v_{t_2}\theta$ for all $t_1,t_2\in \Tm$.

A {\em conditional unification problem} is a finite set of conditional equalities
\begin{equation}\label{CUPG}
A_i=B_i\Rightarrow C_i=D_i, \qquad A_i,B_i,C_i,D_i\in Expr,\;\;i=1,\ldots ,n.
\end{equation}
Its solution, or {\em unifier}, is a comprehensive idempotent ($\theta^2=\theta$)
substitution $\theta\!: \Expr\to \Expr$ such that $A_i\sigma=B_i\theta$ implies
$C_i\theta=D_i\theta$ for $i=1,\ldots ,n$. The conditional unification problem  is
called {\em unifiable} when such a unifier does exist.

The classical (unconditional) first-order  unification is a special case of this
definitions. In our case the main results of the classical unification theory are also
valid. It was established in \cite{KruAPAL} for the first-order conditional unification;
the case of a language with reference constructions of the form $v(t)$ was considered in
\cite{KruJLC}, \cite{KruTCS} where the following statements were proved:
\footnote[3]{The general second-order unification problem is known to be undecidable
\cite{Farm},\cite{Gold}. In our case it is decidable. The problem is more simple because
there is no nested occurrences of the funtion variable $v$ in the language. }

\begin{itemize}
\item The unifiability property for conditional unification problems of the form
(\ref{CUPG}) is decidable.

\item Any unifiable problem of the form (\ref{CUPG}) has a unifier $\theta$ that is {\em
the most general unifier} (m.g.u) in the following {\em weak sense}: any substitution
$\theta'$ that unifies (\ref{CUPG}) has the form $\theta'=\theta\lambda$ for some
substitution $\lambda$. (Note that not every substitution of the form $\theta \lambda $
must unify (\ref{CUPG}).)

\item The m.g.u. of (\ref{CUPG}) can be computed effectively given $A_i,B_i,C_i,D_i$,
$i=1,\ldots,n$.

\item The computation of $\theta$ can be detailed in the following way. Let $V$ be the
set of all variables $v\in Var$ that occur in (\ref{CUPG}). It is possible to compute a
finite substitution $\theta_0$ with $Dom(\theta_0)\subseteq V$ such that
\begin{equation}\label{EXTEDEDSUBST}
z\theta = \left\{
\begin{array}{ll}
z\theta_0, &\mbox{if } z\in V, \\
z, & \mbox{if } z\in (P\cup J)\setminus V, \\
(v_{t\theta_0})\theta_0, &\mbox{if } z=v_t \in v(Tm)\setminus V.
\end{array}
\right .
\end{equation}
We may also assume that $\theta_0$ is {\em conservative}, i.e.
\begin{equation}\label{CONSERVATIVE}
Var(\theta_0)\subseteq V\cup \{v_{t\theta_0}\mid v_{t}\in V\}.
\end{equation}
\end{itemize}

The finite substitution $\theta_0$ (together with the finite set $V$) can be used as a
finite representation of the most general unifier $\theta$. We will call it {\em the
finite part} of $\theta$. It can be computed by the variable elimination method, so if
two conditional unification problems $S$ and $S'$ are unifiable, $S\subseteq S'$, and
$\theta$ is a m.g.u. of $S$ with the finite part $\theta_0$, then it is possible to
choose a m.g.u. $\theta'$ of $S'$ with the finite part $\theta_0'$ for which
$Dom(\theta_0)\subseteq Dom(\theta_0')$. In this case we will write
$\theta\preceq\theta'$. Note that if $\theta\preceq\theta'$ and $Dom(\theta_0)=
Dom(\theta_0')$ then $S'$ has the same unifiers as $S$.

\begin{defin}\rm
Let $S$ be the conditional unification problem (\ref{CUPG}) and $A,B\in\Expr$. We shall
write \ $A=B\mod S$ \ when $A\theta= B\theta$ for every unifier $\theta$ of $S$.
\end{defin}

\begin{lem}[\cite{KruJLC},\cite{KruTCS}]
\label{unifcalc} The relation $A=B\mod S$ \ is decidable.
\end{lem}
\proof The unifiability property of $S$ is decidable. If $S$ is not unifiable then
$A=B\mod S$ holds for every $A,B\in \Expr$. For unifiable $S$ one should restore the
most general unifier $\theta$ of $S$ and test the equality $A\theta= B\theta$. \qed

With a formula of the form $G=\bigwedge_{i=1}^n t_i\!\!:\!F_i$ we associate a
conditional unification problem:
\begin{equation}\label{CUP}
t_i=t_j\Rightarrow F_i=F_j,\quad i,j=1,\ldots ,n.
\end{equation}
We shall write  $A=B\mod \,G$ \ when $A=B\mod S$ \ and $S$ is the conditional
unification problem~(\ref{CUP}).

\section{Referential justification logic $\Jref$}
The idea to express the injectivity of justifications via unification first appeared in
\cite{ArtStr93TR}. Later it was developed in order to axiomatize the single-conclusion
property of arithmetical proof predicates (see \cite{Kru}, \cite{KruAPAL},
\cite{KruJLC}, \cite{KruTCS}). It was used for axiomatization of symbolic models of
single-conclusion proof logics in \cite{KruCSR}, \cite{KruMSb}. The concept of an
injective basic justification model is more general, so we  extend this approach.

We will distinguish between the language of a basic justification model and the language
$L^{v}(P,J)$ that will be used to formulate the properties of the model.

\begin{defin}\label{SEMANTICS}\rm
{\em An interpretation} of the language $L^{v}(P,J)$ in a basic justification model
$M=\langle L(P^0,J^0), *\rangle$ is a comprehensive (infinite) substitution $\sigma$
that maps terms and formulas of the language $L^{v}(P,J)$ into terms and formulas of the
language $L(P^0,J^0)$ respectively, \ $\sigma\!: Tm \to Tm^0$, \ $\sigma\!: Fm \to
Fm^0$. We also require that $v_t\sigma\in (t\sigma)^*$ when $(t\sigma)^*$ is nonempty.
The corresponding validity relation for formulas $F\in Fm$ is defined in the usual way:
$$
\langle \sigma , M\rangle\models F \quad \mbox{iff}\quad (F\sigma)^*=1.
$$
\end{defin}

Referential justification logic $\Jref$ \ in the language $L^{v}(P,J)$ is defined by the
following calculus:

\noindent
\begin{description}
\item[(A0)]  axioms of the classical propositional logic, \\[-9pt]

\item[(A1)]  $s\!:\!(F\rightarrow G)\rightarrow (t\!:\!F\rightarrow [s\cdot
t]\!:\!G)$,\hfill (Application)\\[-9pt]

\item[(A2)]  $\bigwedge\limits_{i=1}^n\,t_i\!:\!F_i\rightarrow (F\leftrightarrow G)$
\quad
if \quad $F=G\,mod\bigwedge\limits_{i=1}^n\,t_i\!:\!F_i$, \hfill (Unification) \\[-4pt]

\item[(A3)] $t\!:\!F \to t\!:\!v_{t}$, \hfill (Assignment)\\[-9pt]

\item[(A4)] $[s\cdot t]\!:\!v_{s\cdot t}\;\to\; s\!:\!(v_{t}\to v_{s\cdot t})\wedge
t\!:\!v_{t}$. \hfill (Sharpness)

\end{description}

\noindent {\bf Inference rule:} $F\rightarrow G,\,F\vdash G$. \hfill  ( Modus ponens)

\medskip
$\Jref$ extends the justification logic ${\sf J}^{-}$. The set of its axioms is
decidable by Lemma \ref{unifcalc}. We will prove that $\Jref$ is sound and complete with
respect to the class of all interpretations in sharp and injective basic justification
models.

Unification axioms (A2) reflect the injective property (see \cite{KruAPAL},
\cite{KruTCS}). Assignment axioms (A3), together with Unification, provide the correct
values for reference variables $v_{t}$ when $t\!:\!F$ is valid (the statement $v_{t}$
restored from $t$ must be equivalent to $F$). The last axiom scheme (A4) makes it
possible to reconstruct logical reasonings given the term structure of justifications.
It means the sharpness property.

\begin{theo} Let $\sigma$ be an interpretation of $L^{v}(P,J)$ in a sharp
injective basic justification model $M=\langle L(P^0,J^0), *\rangle$ and $F\in Fm$. Then
$\Jref\vdash F$ implies $\langle \sigma , M\rangle\models F$.
\end{theo}
\proof It is sufficient to prove that the translations of axioms (A0)-(A4) are valid in
$M$. For (A0), (A1) it follows from the fact that $M$ is a model for ${\sf J}^{-}$.

Case (A2). Suppose that $\langle \sigma , M\rangle\models \bigwedge\limits_{i=1}^n
\,t_i\!:\!F_i$, so
$$
(t_i\sigma)^*=\{F_i\sigma\},\quad i=1,\ldots,n.
$$

There exists a unifier $\theta$ of (\ref{CUP}) such that
\begin{equation}\label{SUBSTEQUIVALENCE}
e_1\sigma=e_2\sigma \Leftrightarrow e_1\theta=e_2\theta
\end{equation}
holds for all expressions $e_1,e_2$ occurring in (A2). Indeed, let $V$ be the finite set
of all variables $v\in Var$ that occur in (A2) and $\sigma_0$ be the restriction of
$\sigma$ to $V$,
$$
z\sigma_0= \left \{
\begin{array}{ll}
z\sigma, & z\in V, \\
z, & z\in Var\setminus V.
\end{array}
\right .
$$
Consider a substitution $\theta_0=\sigma_0\lambda$ where $\lambda$ is an injective
substitution that maps $P^0$ into $(P\setminus V)$ and $J^0$ into $(J\setminus V)$. The
substitution $\theta_0$ maps $\Expr$ into $\Expr$ and is idempotent, but satisfies the
limited comprehension condition ($t_1\theta_0=t_2\theta_0\To
v_{t_1}\theta_0=v_{t_2}\theta_0$) only for terms that occur in (A2). The full-scale
comprehension will be forced by the transformation (\ref{EXTEDEDSUBST}). The
corresponding substitution $\theta$ is comprehensive and idempotent. It coincides with
$\theta_0$ on variables from $V$, so the equivalence (\ref{SUBSTEQUIVALENCE}) follows
from the injectivity of $\lambda$.

We claim that $\theta$ is a unifier of (\ref{CUP}). Indeed,
$$
t_i\theta=t_j\theta \;\To\; (t_i\sigma)^*=(t_j\sigma)^* \;\To\; F_i\sigma=F_j\sigma
\;\To\; F_i\theta=F_j\theta.
$$

But \ $F=G\,mod\bigwedge\limits_{i=1}^n\,t_i\!:\!F_i$ \ implies $F\theta=G\theta$ and
$F\sigma=G\sigma$. Thus, $F$ and $G$ denote the same formula in the language
$L(P^0,J^0)$, so $\langle \sigma , M\rangle\models (F\leftrightarrow G)$.

Case (A3) follows from the definition of the translation. If $\langle \sigma ,
M\rangle\models t\!\!:\! F$ then $v_t\sigma = F\sigma$ because $M$ is injective, so
 $t\!\!:\! F$ and $t\!\!:\!v_t$ denote the same formula in the language
$L(P^0,J^0)$.

Case (A4). Suppose $\langle \sigma , M\rangle\models [s\cdot t]\!:\!v_{s\cdot t}$. Then
$v_{s\cdot t}\sigma \in (s\sigma\cdot t\sigma)^*$. By the sharpness property of $M$,
there exists a formula $F$ such that $F\in (t\sigma)^*$ and $(F\to v_{s\cdot t}\sigma)
\in (s\sigma)^*$. But $v_{t}\sigma\in (t\sigma)^*$ because
 $(t\sigma)^*$ is nonempty, so $F=v_{t}\sigma$ by the injective property of $M$.
 Thus, $\langle \sigma , M\rangle\models s\!:\!(v_{t}\to v_{s\cdot
t})\wedge t\!:\!v_{t}$. \qed

\section{Completeness}
\begin{theo}\label{COMPLETENESS}
Let $\Jref\not\vdash F$. There exists an interpretation $\sigma$ of the language
$L^{v}(P,J)$ in a sharp injective basic justification model $M$ such that $\langle
\sigma,M\rangle\not\models F$.
\end{theo}

The completeness proof is based on the saturation procedure from \cite{KruJLC},
\cite{KruTCS} where its general form for languages with reference constructions is
developed. We will use a simplified version that fits the language $L^{v}(P,J)$.

Let $(\theta, \Gamma,\Delta)$ be the global data structure, where
$\theta\!:\Expr\to\Expr$ is a substitution\footnote[4]{$\theta$ is  an infinite
substitution of the form (\ref{EXTEDEDSUBST}). We store the finite part of it.} and
$\Gamma,\Delta\subset Fm$ are finite sets of formulas. The saturation is a
nondeterministic procedure that starts from a formula $F\in Fm$. It initializes the data
structure: $\theta := id$, $\Gamma := \emptyset$, $\Delta := \{\bot,F\}$. Then it
applies repeatedly the following blocks of instructions:

\begin{enumerate}
\item For every $X\to Y \in \Gamma$ that has not been discharged by the rule 1 before
nondeterministically add $Y$ to $\Gamma$ or add $X$ to $\Delta$. Discharge $X\to Y$ and
all its descendants (its substitutional instances  that will be added to $\Gamma$ by
block \ref{SUBSINST} later). For every $X\to Y \in \Delta$ add $X$ to $\Gamma$ and add
$Y$ to $\Delta$. Repeat these actions until $\Gamma,\Delta$ will not change. If
$\Gamma\cap\Delta \not=\emptyset$ then terminate with failure else go to 2.

\item For every $t\!:\!X\in \Gamma$ add $t\!:\!v_t$ to $\Gamma$. For every  term $t$
that occurs in some formula from $\Gamma\cup\Delta$ do: if $t\theta\!:\!v_{t\theta}\in
\Gamma$ add $t\!:\!v_t$ to $\Gamma$. For every $[s\cdot t]\!:\!X\in \Gamma$ also add
$s\!:\!(v_t\to X)$ and $t\!:\!v_t$ to $\Gamma$. For every pair $s\!:\!(X\to
Y),\,t\!:\!X\in\Gamma$ do: if the term $s\cdot t$ occurs in some formula from
$\Gamma\cup\Delta$ then add $[s\cdot t]\!:\! Y$ to $\Gamma$. Repeat these actions until
$\Gamma$ will not change. If $\Gamma\cap\Delta \not=\emptyset$ then terminate with
failure else go to 3.

\item\label{SUBSINST} Combine a formula $t_1\!:\!F_1\wedge\ldots\wedge t_n\!:\!F_n$
where $t_i\!:\!F_i$, $i=1,\ldots,n$ are all formulas of the form $t\!:\!X$ from
$\Gamma$. Test the corresponding unification problem (\ref{CUP}) for unifiability. If it
is not unifiable then terminate with failure. If it is unifiable then compute an m.g.u.
$\theta'\succeq\theta$ of (\ref{CUP}) and update $\Gamma:=\Gamma\cup \Gamma\theta'$,
$\Delta:=\Delta\cup\Delta\theta'$. If $\Gamma\cap\Delta \not=\emptyset$ then terminate
with failure. Otherwise compare the finite parts $\theta_0'$ and $\theta_0$. If
$Dom(\theta_0')= Dom(\theta_0)$ then set $\theta:=\theta'$ and terminate with success;
else update $\theta:=\theta'$ and go to 1.

\end{enumerate}

Consider a computation of the saturation procedure. Any action in it that changes the
data structure $(\theta, \Gamma,\Delta)$ will be called a {\em saturation step}. There
are steps of type 1, 2 or 3 depending on the block involved.

\begin{lem}\label{TERMINATES} Every computation of the saturation procedure terminates.
\end{lem}
\proof Consider a computation starting from $F$. Suppose that it does not terminate with
failure. It is sufficient to prove that it contains a finite number of steps.

Let
$$
V^i=V^i_1\cup V^i_2, \quad V^i_1\subset (P\cup J),\quad V^i_2\subset v(Tm)
$$
be the set of all variables occurring in $\Gamma\cup\Delta$  and $T^i$ be the set of all
terms occurring in $\Gamma\cup\Delta$ \ at some state $i$ of the computation.

The computation does not change the set $V^i_1$ because all substitutions constructed by
steps of type 3 are conservative (see (\ref{CONSERVATIVE})). All variables of a term
$t\in T^i$ belong to $V^i_1$. Steps of types 1,2 do not change the set $T^i$. Steps of
type 3 may extend the set $T^i$ by terms of the form $t\theta'$, $t\in T^i$, but the
choice of $\theta'\succeq \theta$ together with the idempotency of m.g.u.'s imply that
sets $T^i$ will stabilize after some steps too. One more iteration after it will
stabilize the set $V_2$. Consider the part of the computation after it.

Consider two consecutive iterations of blocks 1-3. Suppose that at the start of the
second one there exists a formula $X\to Y\in \Gamma\cup\Delta$ that is not discharged.
It is obtained at the previous iteration from some variable $p\in \Gamma\cup\Delta$ by
substitution $\theta$  executed by block 3,
$$
X\to Y= p\theta,\quad  p\in P\cup v(Tm).
$$
Formula $X\to Y$ and all its descendants will be discharged at the second iteration by
block~1. It means that $p$ will be never used in this role later because later the
substitution will be updated as $\theta'=\theta\lambda$ and
$p\,\theta'=p\,\theta\lambda= p\,\theta^2\lambda=(X\to Y)\theta'$, so $p\,\theta'$ will
be a descendant of $X\to Y$ and must be already discharged. Thus, the number of
iterations with active steps of type 1 does not exceed the maximal cardinality of sets
$V^i$ plus one. Two iterations after the last active step of type~1 will stabilize the
conditional unification problem (\ref{CUP}) extracted from $\Gamma$ and terminate the
computation with success. \qed

Let the initial formula $F$ be fixed.  All computations starting from $F$ form a
saturation tree. It has no infinite paths by Lemma \ref{TERMINATES}. Its brunching is
bounded, so the saturation tree is finite.

\begin{lem}\label{PROVABLE}
If all computations starting from $F$ terminate with failure then $\Jref\vdash F$.
\end{lem}

\proof Consider a node of the saturation tree. Let  $\Gamma,\Delta$ be the contents of
the data structure at that node. One can establish  by the straightforward induction on
the depth of the node that $\Jref\vdash \bigwedge\Gamma\to\bigvee\Delta$. For the root
node it implies $\Jref\vdash F$. \qed

\medskip\noindent
{\it Proof of Theorem \ref{COMPLETENESS}.} Suppose $\Jref\not\vdash F$. By Lemma
\ref{PROVABLE}, there exists a successful computation of the saturation procedure
starting from $F$. Let $(\theta, \Gamma,\Delta)$ be the resulting contents of the data
structure, $\Expr'=\{e\theta\mid e\in\Expr\}$,
$$
Var'=Var\cap\Expr',\quad Tm'=Tm\cap\Expr',\quad Fm'=Fm\cap\Expr',
$$
$$
\Gamma'=\Gamma\cap\Expr',\qquad \Delta'=\Delta\cap\Expr'.
$$

The substitution $\theta$ is idempotent, so the set $\Expr'$ consists of all fixed
points of $\theta$. For every term $t\in Tm'$ the set $Fm'$ contains at most one formula
of the form $t\!:\!X$ because $\theta$ is comprehensive.

Completion. We construct the set $\Gamma''\supseteq\Gamma'$,
$\Gamma''\cap\Delta'=\emptyset$, and the substitution $\lambda\!:\Expr'\to \Expr'$ as
follows. Consider a pair of formulas $s\!:\!(X\to Y)$, $t\!:\! X\in \Gamma'$ such that
$[s\cdot t]\!:\!Y\not\in \Gamma'$. By the restriction from saturation block 2, $[s\cdot
t]\!:\!Y\not\in \Delta'$, the variable $v_{s\cdot t}$ does not occur in formulas from
$\Gamma'\cup\Delta'$ and $v_{s\cdot t}\in Var'$. Add $[s\cdot t]\!:\!Y$ to $\Gamma'$ and
set $v_{s\cdot t}\lambda :=Y$. Note that the set of all variables occurring in formulas
from $\Gamma'\cup\Delta'$ remains unchanged. Repeat this step until $\Gamma'$ will not
change and define $\Gamma''$ as the least fixed point of it.

The substitution $\lambda$ defined by this process is idempotent, $Dom(\lambda)\subset
v(Tm')$, $Var(\lambda)\subset Var'$ and $X\lambda=X$ for $X\in \Gamma'\cup\Delta'$. Let
$$P^0=\{p\in Var'\mid p\lambda=p\},\qquad J^0=Var'\cap J.$$
Consider the language $L(P^0,J^0)$ with  the interpretation $*$ defined by $\Gamma''$:
$$
p^*=1 \Leftrightarrow p\in\Gamma'' \quad\mbox{ for }p\in P^0,
$$
$$
t^*=\{X\mid t\!:\!X\in \Gamma''\} \quad\mbox{ for }t\in Tm^0.
$$
By the construction, it is a basic justification model $M$ that is sharp and injective.
The sharpness condition is forced by saturation block 2 and the completion procedure.
The model is injective because for each $t$ the set $\Gamma'$  contains at most one
formula of the form $t\!:\!X$ and the completion procedure preserves this property.

\begin{lem}[Truth lemma]
If $G\in\Gamma''$ then $G^*=1$, if $G\in\Delta'$ then $G^*=0$.
\end{lem}
\proof Straightforward induction on the complexity of $G$. Note that
$\Gamma''\cap\Delta'=\emptyset$. If $G$ is atomic or has the form $t\!:\!X$ then the
statement follows from the definition of *. For $G$ of the form $X\to Y$ it is forced by
saturation block 1. In this case $G\in\Gamma'\cup\Delta'$, so it will be discharged by
block 1 at some step. \qed

The substitution $\sigma=\theta\lambda$ is an interpretation of the language
$L^{v}(P,J)$ in $M$. Indeed, it is idempotent because $\theta$ and $\lambda$ are
idempotent and $Var(\lambda)\subset Var'$. It is comprehensive because $\theta$ is
comprehensive and $Dom(\lambda)\subseteq v(Tm')$. As a consequence, the equality
$v_t\sigma=v_{t\sigma}\sigma$ holds for each $t\in Tm$.

Suppose $(t\sigma)^*\not=\emptyset$ for some $t\in Tm$. Then $t\sigma=t\theta=t'$,
$(t')^*=\{X'\}$ and $t'\!:\!X'\in \Gamma''$ for some $t'\in Tm'$, $X'\in Fm'$. If
$t'\!:\!X'\in \Gamma'$ then, by saturation block~2, \ $t'\!:\!v_{t'}\in \Gamma'$, and
$v_{t'}\theta = X'\theta= X'$ by saturation block~3. But in this case
$v_t\sigma=v_{t'}\theta$ because $X'\lambda=X'$. If \ $t'\!:\!X'\in
\Gamma''\setminus\Gamma'$ then $v_{t'}\lambda=F'$ by the definition of $\lambda$ and
$v_t\sigma=v_{t'}\lambda$. In both cases \ $v_t\sigma=X'\in (t\sigma)^*$.

We have $F\sigma=F\theta\in \Delta'$. By Truth lemma, $(F\sigma)^*=0$, so $\langle
\sigma,M\rangle\not\models F$. \qed

\begin{cor}
The logic $\Jref$ is decidable.
\end{cor}
\proof $\Jref\vdash F$ iff all computations of the saturation procedure starting from
$F$ terminate with failure. The saturation tree is finite and can be restored from $F$.
\qed

{\bf Comment.}  Basic justification models that are injective but not necessarily sharp
can be axiomatized in the language $L(P,J)$ without function variable $v$ by axioms
(A0)-(A3). The definition of a unifier used in (A3) should be simplified by omitting the
comprehension condition and all other items that involve expressions of the form $v_t$.
The corresponding justification logic is also decidable.

\section*{Acknowledgements}
I would like to thank Sergei Artemov who attracts my attention to the problem.

\end{document}